\documentclass[aps,prd,twocolumn,twoside,superscriptaddress,floatfix, nofootinbib]{revtex4-1}

\usepackage{graphicx}
\usepackage{ifthen,amsmath,amssymb, bm}
\usepackage{graphicx}
\usepackage[dvipsnames]{xcolor}
\usepackage{hyperref}
\usepackage{float}
\usepackage{aas_macros}
\usepackage{caption}
\usepackage{subcaption}
\usepackage[export]{adjustbox}
\usepackage{bookmark}

\def\mpcoh{\,h^{-1}{\rm Mpc}} 
 
\def\hompc{\,h\,{\rm Mpc}^{-1}}

\begin{document}

\title{Predicting the 21 cm field with a Hybrid Effective Field Theory approach}

\author{Danial Baradaran}
\affiliation{Berkeley Center for Cosmological Physics, Department of Physics, University of California, Berkeley, CA 94720, USA}

\author{Boryana Hadzhiyska}
\email{boryanah@berkeley.edu}
\affiliation{Physics Division, Lawrence Berkeley National Laboratory, Berkeley, CA 94720, USA}
\affiliation{Berkeley Center for Cosmological Physics, Department of Physics, University of California, Berkeley, CA 94720, USA}

\author{Martin White}
\affiliation{Physics Division, Lawrence Berkeley National Laboratory, Berkeley, CA 94720, USA}
\affiliation{Berkeley Center for Cosmological Physics, Department of Physics, University of California, Berkeley, CA 94720, USA}

\author{Noah Sailer}
\affiliation{Berkeley Center for Cosmological Physics, Department of Physics, University of California, Berkeley, CA 94720, USA}
\affiliation{Physics Division, Lawrence Berkeley National Laboratory, Berkeley, CA 94720, USA}

\begin{abstract}

A detection of the 21 cm signal can provide a unique window of opportunity for uncovering complex astrophysical phenomena at the epoch of reionization and placing constraints on cosmology at high redshifts, which are usually elusive to large-scale structure surveys. In this work, we provide a theoretical model based on a quadratic bias expansion capable of recovering the 21 cm power spectrum with high accuracy sufficient for upcoming ground-based radio interferometer experiments. In particular, we develop a hybrid effective field theory (HEFT) model in redshift space that leverages the accuracy of $N$-body simulations with the predictive power of analytical bias expansion models, and test it against the Thesan suite of radiative transfer hydrodynamical simulations. We make predictions of the 21 cm brightness temperature field at several distinct redshifts, ranging between $z = 6.5$ and 11, thus probing a large fraction of the reionization history of the Universe ($x_{\rm HI} = 0.3 \sim 0.9$), and compare our model to the `true' 21 cm brightness in terms of the correlation coefficient, power spectrum and modeling error. We find percent-level agreement at large and intermediate scales, $k \lesssim 0.5 h/{\rm Mpc}$, and favorable behavior down to small scales, $k \sim 1 h/{\rm Mpc}$, outperforming pure perturbation-theory-based models. To put our findings into context, we show that even in the absence of any foreground contamination the thermal noise of a futuristic HERA-like experiment is comparable with the theoretical uncertainty in our model in the allowed `wedge' of observations, providing further evidence in support of using HEFT-based models to approximate a range of cosmological observables.

\end{abstract}
\maketitle



\section{Introduction} 
\label{sec:intro}


Detecting the hyperfine transition of neutral hydrogen in the form of emitted radiation at a rest wavelength of 21 cm offers a viable path for mapping out structure in the Universe at high redshifts, which are otherwise hard to probe due to the small number of detectable luminous sources. Currently, most of the cosmological information used to test our theories about the Universe comes from either large-scale structure (LSS) surveys at low redshifts $z \lesssim 3$ or the cosmic microwave background (CMB), emitted at $z \approx 1100$ during the epoch of recombination. As such, much of the formation history of the Universe at intermediate redshifts remains unexplored. The ideal tracer of the period spanning between the epoch of recombination and the epoch of reionization (EoR) is the diffuse neutral hydrogen gas, detectable in the form of 21 cm emission. 

The EoR refers to the period during which the first sources of (ionizing) UV radiation turned on and ionized the neutral hydrogen in the Universe \citep{2016ARA&A..54..313M}. In addition to providing avenues for testing central assumptions in cosmology, studying the EoR is invaluable to understanding the formation of galaxies and other complex astrophysical phenomena.
As indicated by CMB and Ly$\alpha$ forest measurements, the mean redshift of reionization is $z \approx 7 - 8$ \citep{2016A&A...596A.108P}, and it is mostly complete by $z \sim 6$ \citep{2015MNRAS.447..499M}. Improvements in the measurements of the reionization process could reveal information about the properties of the ionizing sources \citep{2006MNRAS.365..115F,2007MNRAS.377.1043M}, the abundance of small-scale structures that absorb many of the ionizing photons \citep{2006MNRAS.366..689C,2009MNRAS.394..960C,2008ApJ...682...14F}, and overall help us to better grasp the process of galaxy formation.

Several experiments are already well on their way to directly image the neutral hydrogen and map the 21 cm signal from the EoR. A number of them attempt to do so at the level of the mean global signal (monopole) such as EDGES \citep{2017ApJ...835...49M}, LEDA \citep{2018MNRAS.478.4193P}, PRIZM \citep{2019JAI.....850004P}, and SARAS \citep{2018ExA....45..269S}, while others are after detecting the fluctuations in the signal such as PAPER \citep{2010AJ....139.1468P}, the MWA \citep{2013PASA...30....7T,2013PASA...30...31B}, LOFAR \citep{2013A&A...556A...2V}, HERA \citep{2017PASP..129d5001D}, and eventually the Square Kilometre Array (SKA) \citep{2020PASA...37....2W}. Additionally, efforts such as CHIME \citep{2014SPIE.9145E..22B}, HIRAX \citep{2016SPIE.9906E..5XN}, and CHORD \citep{2019clrp.2020...28V} attempt to trace the post-reionization history of the Universe by adopting intensity mapping techniques. Once a detection of the redshifted 21 cm signal is confirmed, the challenge is to robustly extract information about the EoR and provide constraints on our cosmological model.

To allow for a reliable interpretation of the data, a number of numerical methods have been developed, including semi-analytic reionization models such as 21CMFAST \citep{2011MNRAS.411..955M} and SimFast21 \citep{2010MNRAS.406.2421S}. The advantage of these methods is that they are computationally inexpensive and can therefore be scaled up to cosmological volumes. In these models, ionized regions are typically `painted' on top of a realization of the matter density distribution at the locations where sources are likely to reside \citep{2005MNRAS.363.1031F,2012ApJ...747..126A,2014MNRAS.440.1662S}. While some of these prescriptions can also incorporate models for the photon sinks that retard reionization, they do not directly resolve them and as such are not as faithful on small scales.

The most reliable method for simulating the reionization process is through the use of radiative transfer simulations \citep{1997ApJ...486..581G,2003MNRAS.343.1101C,2006MNRAS.369.1625I,2007ApJ...671....1T,2014ApJ...793...30G,2017MNRAS.466..960P}, which are much more computationally expensive and are therefore often limited to smaller box sizes ($\sim$100 Mpc) \citep{2005ApJ...624L..65B}. The challenge is that these simulations need to have a sufficiently large number of resolution elements in order to capture the smallest intergalactic
structures that act as photon sinks \citep{2000ApJ...530....1M,2000ApJ...542..535G,2004MNRAS.348..753S}, and at the same time have a sufficiently large volume to provide a representative sample of the largest structures found in the Universe. Nonethlesess, these simulations are prohibitively expensive to survey the vast parameter space
of potential source models. However, perturbative models, inspired by Effective Field Theory (EFT) \citep[e.g.,][]{2005MNRAS.363.1031F,2007ApJ...654...12Z,2011MNRAS.414..727Z}, can potentially provide a fast way of analyzing 21 cm observables and thus, elucidate many astrophysical and cosmological conundrums.

Until recently, it was believed that the 21 cm signal cannot be treated perturbatively in the regime probed by 21 cm experiments due to the presence of large ionized bubbles. 
However, developments in the last few years \citep{2007MNRAS.375..324Z,2015PhRvD..91h3015M,2018JCAP...10..016M,2022PhRvD.106l3506Q,2022JCAP...10..007S} suggest otherwise: the 21cm signal is likely perturbative on many
of the scales that future interferometric instruments will be sensitive to. On these scales, the signal is modeled in terms of `bias' parameters, which connect it to the underlying matter density field and which can be tied to physical quantities relevant to the physics of reionization. 

Apart from being able to model the 21 cm field in real space, it is of utmost importance that we develop techniques that work in redshift space as well, where the 21 cm field is affected by the presence of redshift space distortions (RSDs). These contributions to the observed redshift result from the peculiar velocities of the neutral hydrogen along the line-of-sight (LOS) and as such, make it difficult to disentangle the `real-space' distribution of the field from its intrinsic motion along the LOS. Interestingly, future experiments such as HERA will predominantly observe modes that are nearly parallel to the LOS for three main reasons: i) the instrument needs to be shielded from 
ground radio sources; ii) obtaining high spectral resolution is easier than obtaining high angular resolution; iii)  modes parallel to the LOS are less affected by foreground contamination due to the chromatic instrument response, which even in the case of spectrally smooth foregrounds induces an unsmooth ringing that is mostly confined to the low $k_\parallel$ and high $k_\perp$ regime 
\citep[see e.g.,][]{1997ApJ...486..581G,2003MNRAS.343.1101C,2015aska.confE...1K}. This underscores the importance of developing redshift-space methods such as the one proposed in this work.

In this paper, we develop a method originally from cosmological perturbation theory for modeling the 21 cm signal in redshift space. 
Our method provides a viable path towards extracting the cosmological information about the composition of the Universe, encapsulated in the large-scale modes, and the astrophysical information about the morphology of reionization, encoded in the small-scale modes. Since the 21 cm field obeys basic symmetries such as rotational invariance and the equivalence principle, we can write it as an expansion in powers of the density field, tidal field and low-order derivatives on scales larger than the typical size of the ionized bubbles \citep{2009JCAP...08..020M,2018JCAP...10..016M,2018PhR...733....1D,2022PhRvD.106l3506Q}. To enable us to work to smaller scales, we combine the power of such a symmetries-based bias expansion and the accuracy of $N$-body simulations in the non-linear regime. The free (bias) parameters of our theory can be linked to valuable astrophysical quantities and can reveal the answers to questions such as when reionization occurred, what fraction of the Universe is neutral, what the bias of the sources is, and what the characteristic bubble sizes are.

This paper is organized as follows. In Section~\ref{sec:methods}, we provide details about our hybrid perturbative model and describe the radiative transfer simulation Thesan-1, against which we test our model. In Section~\ref{sec:results}, we present our main findings in terms of the two-point correlation function and the relative modeling error, putting it in the context of the sensitivity of future experiments. Finally, we summarize our main conclusions in Section~\ref{sec:conclusions}.

\section{Methods}
\label{sec:methods}

In this section, we describe the radiative transfer simulations adopted in this study and our theoretical model for approximating the 21 cm field.

\subsection{Thesan simulations}

We test the theoretical model developed in this work via the high-resolution radiation-magneto-hydrodynamical simulation suite, Thesan. This suite is designed to faithfully reproduce both the complex physics of reionization as well as galaxy formation at high redshifts. The comoving box size of these simulations is $\approx 65\,h^{-1}$Mpc, and they are sufficiently well-resolved to capture the formation of atomic cooling halos, which are the smallest structures significantly contributing to the process of reionization. The galaxy formation model is built on top of the well-tried and tested IllustrisTNG model \citep{2017MNRAS.465.3291W,2018MNRAS.473.4077P}, whereas the initial conditions are produced according to Ref. \citep{2016MNRAS.462L...1A}. The simulations are run using the radiative transfer version of the Arepo code, Arepo-RT \citep{2010MNRAS.401..791S,2019MNRAS.485..117K}. Radiation-magnetohydrodynamics equations are solved on a mesh, with mesh-generating points that approximately follow the gas flow via their Voronoi tessellation. 

The Thesan simulations have been shown to reproduce a number of observables of the reionization history of the Universe, the evolution of the temperature of the intergalactic medium (IGM), the optical depth to the CMB, the UV luminosity function at $z \geq 6$ \citep{2022MNRAS.514.3857K}, the photo-ionization rate, the mean-free-path of ionizing photons, the IGM opacity, and the temperature-density relation \citep{2022MNRAS.512.4909G}. In this work, we employ Thesan-1, the highest-resolution simulation, which contains 2100$^3$ dark matter particles with mass $3.12 \times 10^6 \ M_\odot$ and 2100$^3$ gas particles of mass $5.82 \times 10^5 \ M_\odot$, and its dark-matter-only ($N$-body) counterpart Thesan-Dark-1, which contains 2100$^3$ dark matter particles with mass $3.702 \times 10^6 \ M_\odot$. The cosmology of the suite is identical to that of IllustrisTNG: $\Omega_m = 0.3089$, $\Omega_b = 0.0486$, $\Omega_\Lambda = 0.6911$, $h = 0.6774$, $\sigma_8 = 0.8159$, $n_s = 0.9667$. 

\begin{figure}
\includegraphics[width=\columnwidth]{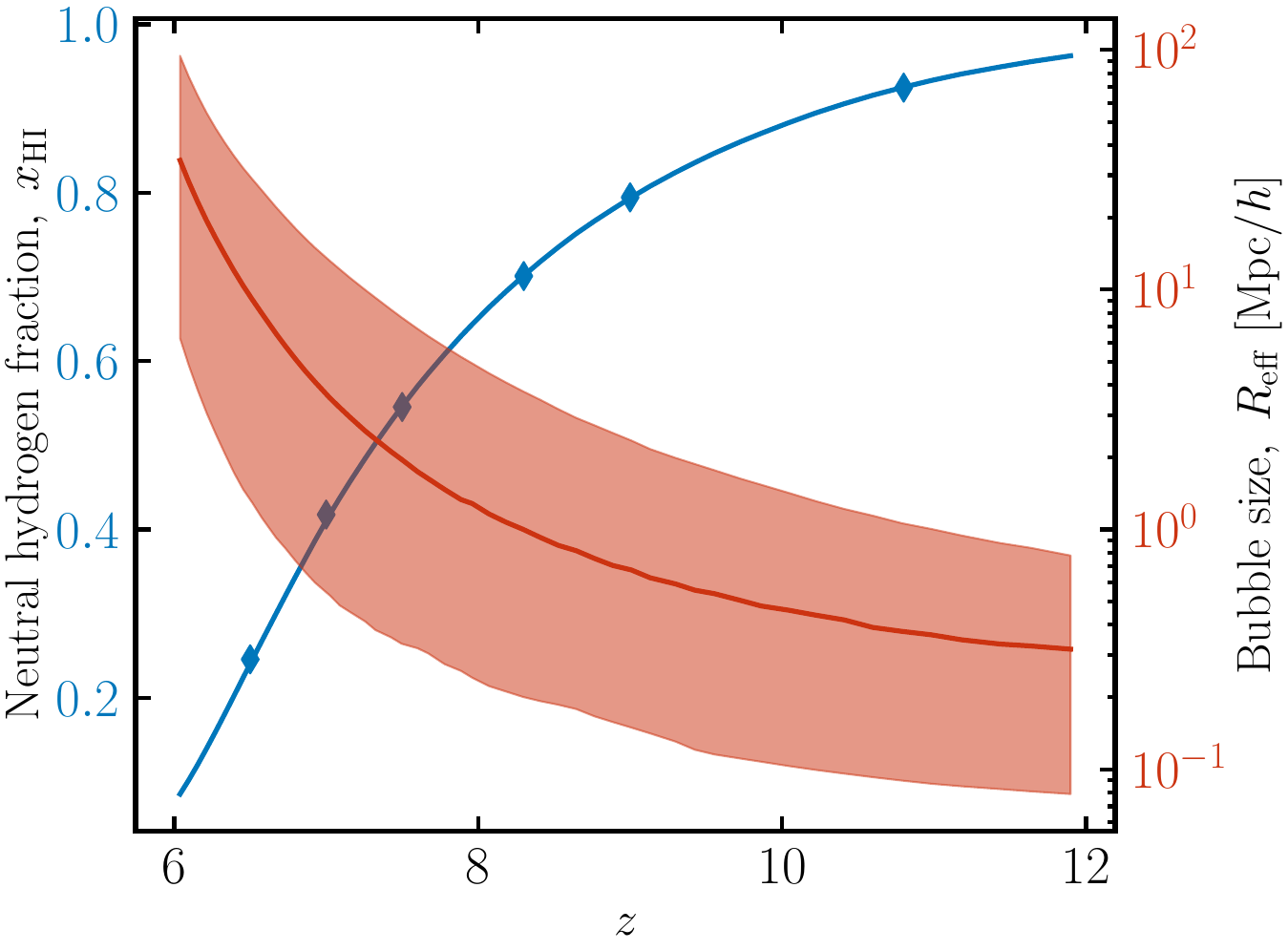}
\caption{Reionization history of the Thesan-1 simulation, shown as the mean neutral hydrogen fraction as a function of redshift. The blue diamond markers indicate the redshifts at which we test the perturbative method developed in this work. The right axis shows the characteristic bubble size (median and 16th to 84th percentile band), identified through the mean-free path analysis of Ref.~\citep{2024MNRAS.tmp.1328N}. Towards the end of reionization $x_{\rm HI} < 0.5$, the bubbles expand exponentially, increasing the scale range affected by non-local physics.}
\label{fig:mean_ion}
\end{figure}

In this work, we utilize outputs at several different redshifts: $z = 10.8$, 9, 8.3, 7.5, 7, and 6.5, with neutral hydrogen fraction of $x_{\rm HI} = 0.92$, 0.8, 0.71, 0.57, 0.45 and 0.3, respectively, thus sampling a large fraction of the reionization history of the Universe. We show the reionization history predicted by the Thesan-1 simulation in Fig.~\ref{fig:mean_ion}. Consistent with CMB experiments, the mid-point in reionization according to Thesan-1 occurs around $z\simeq 7-8$. Most of the redshifts at which we probe our theoretical model are in the first half of reionization, where the bubble sizes are smaller, and we expect the 21 cm field to be more perturbative. We also display the characteristic bubble size in the same figure, identified through the mean-free path analysis of Ref.~\citep{2024MNRAS.tmp.1328N}. Towards the end of reionization $x_{\rm HI} < 0.5$, the bubbles expand exponentially, increasing the scale range affected by non-local physics. Identifying the scale of non-locality is crucial for perturbative models, as it indicates the regime in which we expect the model to break.

\subsection{Theoretical model}
\label{sec:theory}

In this section, we summarize our approach for modeling the 21 cm signal, which is inspired by the treatment of biased tracers in Lagrangian Perturbation Theory (LPT). For a review of perturbation theory, see  Ref.~\cite{2002PhR...367....1B,2018PhR...733....1D,2020moco.book.....D}. 

In the Lagrangian picture, we work with infinitesimal fluid elements labeled by their initial (Lagrangian) positions $\bm{q}$. Their dynamics are encoded in a displacement vector $\bm{\Psi}(\bm{q},\eta)$, sourced by the gravitational potential and defined such that the Eulerian (comoving) positions $\bm{x}$ of the fluid element at some conformal time $\eta$ is $\bm{x}(\bm{q},\eta) = \bm{q} + \bm{\Psi}(\bm{q},\eta)$ \cite{2015JCAP...09..014V,2014JCAP...05..022P}. The 3D distribution of cosmological observables such as the galaxy field or the 21cm signal is, in general, determined by complex small-scale astrophysical processes and their response to the large-scale matter distribution \citep{2018PhR...733....1D}.

We can approximate the behavior of any field obeying simple symmetries such as rotation invariance and the equivalence principle in terms of the matter density, velocity gradients and tidal fields in a neighborhood around its trajectory, i.e., within the scale of `nonlocality'. In the case of galaxies, this roughly corresponds to the size of the Lagrangian halo radius, since the dominant physical process is the gravitational collapse of the Lagrangian patch into a halo. On the other hand, the 21cm signal is sensitive to potentially larger nonlocalities resulting from the mean free path of ionizing photons, which can travel significant distances in the absence of absorption. In that case, the size of the nonlocality is roughly set by the size of the ionized bubbles. Thus, at higher redshifts, when the bubbles are smaller, we can use perturbation theory to model the signal over a wider range of scales \cite{2018JCAP...10..016M}.

We can thus write the dependence of the 21cm signal along its trajectory as an expansion to second order in the initial conditions adopting the Lagrangian EFT framework \cite{2009JCAP...08..020M,2015JCAP...11..007S,2014JCAP...08..056A,2016JCAP...12..007V}:
\begin{align}
\label{eq:bias_expansion}
  F(\bm{q}) =& 1 + b_1 \delta_L(\bm{q}) + b_2 \big(\delta_L^2(\bm{q})-\langle\delta_L^2\rangle\big) \\ \nonumber &+ b_s \big(s_L^2(\bm{q})-\langle s_L^2\rangle \big) + b_\nabla \nabla^2 \delta_L(\bm{q}) , \end{align}
where $b_1$, $b_2$, $b_s$ and $b_\nabla$ are free bias parameters and $s_L^2 = s_{ij} s^{ij}$ is the shear/tidal field, with $s_{ij}\equiv ( \partial_i \partial_i/\partial^2 - \delta_{ij}/3 )\ \delta_L$. Note that the functional $F(\bm{q})$ can only depend on scalar combinations of the density field and the tidal field. Nonlocality effects are handled by performing a Taylor expansion of the $\delta_L$ field around $\bm{q}$. To lowest order, this calls for the inclusion of $\nabla^2\delta_L$. We note that on nonlocal scales, i.e., smaller than the bubble size, the Lagrangian approximation breaks down. 

The functional can then be advected to the real-space (Eulerian) position $\bm{x}$ \cite{2008PhRvD..77f3530M}:
\begin{equation}
    1 + \delta_{\rm 21}(\bm{x}) 
   = \int d^3\bm{q}\,F(\bm{q})\, \delta^D(\bm{x}-\bm{q}-\bm{\Psi}(\bm{q}))
\label{eqn:delta21}
\end{equation}
Thus, the 21cm clustering signal contains a dynamics piece, $\bm{\Psi}$, as well as a piece depending on the initial conditions, $F(\bm{q})$. 

\begin{figure*}
\includegraphics[width=2\columnwidth]{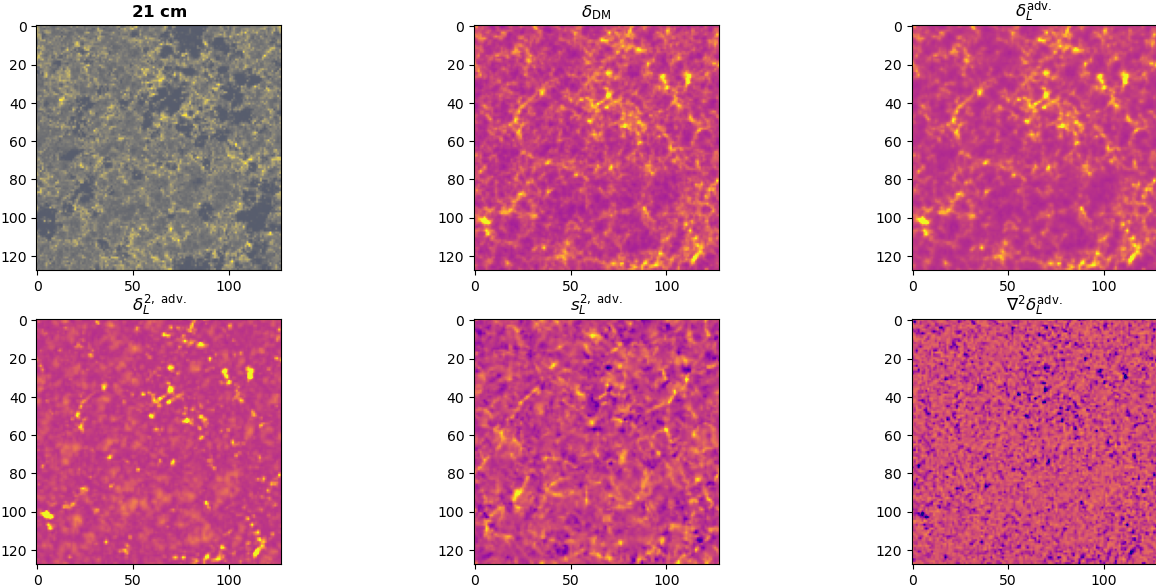}
\caption{21 cm field (top left) and Lagrangian operators (remaining five panels) at $z = 9$ at the fiducial resolution of this analysis, $N_{\rm mesh} = 128^3$. The Lagrangian operator fields are obtain by advecting the initial condition density fields, $\delta_L$, $\delta^2_L$, $s^2_L$ and $\nabla^2 \delta_L$, via a time-evolved $N$-body simulation. Visible in the 21 cm field are the bubble nucleation sites (dark spots) and the regions of high HI density (lighter spots). Our theoretical model of the 21 cm field, described in Section~\ref{sec:theory}, is constructed by linear combinations of the five advected fields. We note this is an $xy$ cross-section of the redshfit-space 21 cm field with LOS along the $z$ axis.
}
\label{fig:fields}
\end{figure*}

\begin{figure*}
\includegraphics[width=2\columnwidth]{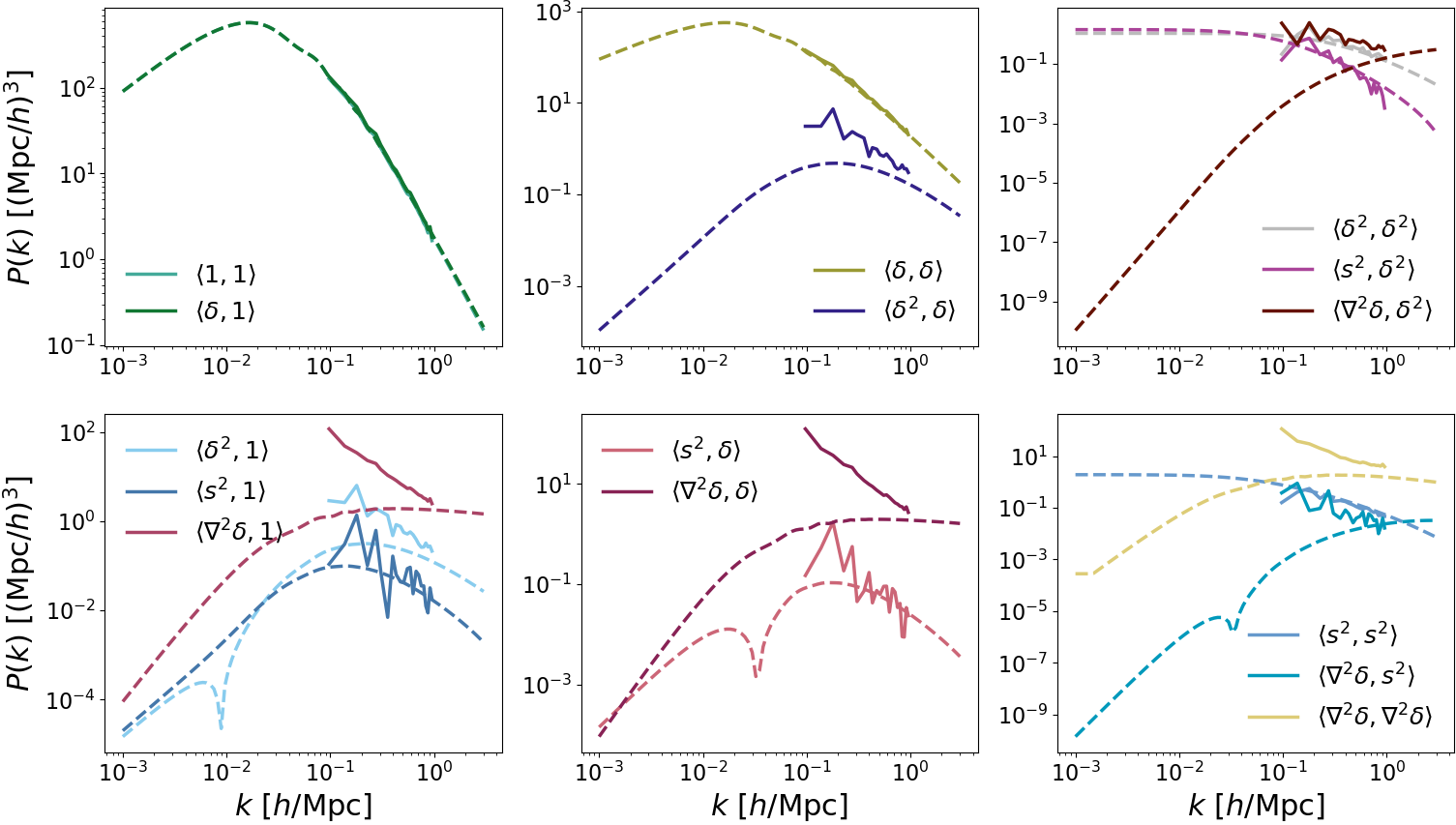}
\caption{Cross- power spectra of all Lagrangian operators in real space, 15 unique combinations in total. Shown here is the comparison between the 2nd order LPT theoretical prediction (computed via \texttt{velocileptors}, dashed lines) and the numerical result (solid lines) obtained using the dark-matter-only counterpart of Thesan-1. We see good agreement between the two for all combinations except those involving the $\nabla^2 \delta$ due to numerical instabilities on small scales. These differences are well understood and irrelevant to our later analysis, as they can be reconciled via a renormalization. As the size of the simulation box is small, we see a significant amount of noise especially in the squared operators.
}
\label{fig:comparison}
\end{figure*}

To account for the fact that we observe the signal in redshift space, $\bm{s}$, and thus pick up contributions in the form of redshift-space distortions (RSD) due to the peculiar velocity along the LOS, $\bm{u} = \hat{\bm{n}}\ (\hat{\bm{n}} \cdot \bm{v}_{\rm pec}) / \mathcal{H}$, we write the redshift-space signal as \cite{2008PhRvD..78j9901M,2019JCAP...03..007V,2021JCAP...03..100C}:
\begin{equation}
    1 + \delta_{\rm 21}(\bm{s}) 
   = \int d^3\bm{q}\,F(\bm{q})\, \delta^D(\bm{s}-\bm{q}-\bm{\Psi}_s(\bm{q})),
\label{eqn:delta21_rsd}
\end{equation}
where $\bm{\Psi}_s$ is the redshift-space displacement field, defined as:
\begin{equation}
    \bm{\Psi}_s(\bm{q}) = \bm{\Psi} + \bm{u} .
\end{equation}

In this work, instead of adopting the usual LPT steps of perturbatively expanding the displacement fields $\bm{\Psi}$ and $\bm{\Psi}_s$, we use an $N$-body simulation as an accurate solver of the dynamics of the matter field. This method has recently been referred to in the literature as Hybrid Effective Field Theory (HEFT) \citep{2020MNRAS.492.5754M,2021JCAP...09..020H,2021MNRAS.505.1422K}. While in real space, this is a trivial operation as described below, in redshift space, we need to be a bit more careful, as taking directly the peculiar velocity field $\bm{u}$ and adding it to the displacement field $\bm{\Psi}$ can incur unwanted contributions from the motions of particles on small scales, also known as Finger-of-God (FoG) effects, which can affect observables such as the power spectrum at the scales of cosmological interest. This is exacerbated by the fact that we use the highest-resolution simulation with 2100$^3$ particles, for which such FoG effects are inevitable. To mitigate this effect, we low-pass filter the LOS velocity field $\bm{u}$ in Fourier space with the following kernel:
\begin{equation}
    \beta(k) = 1 - \tanh\left[\frac{k - k_0}{\Delta k}\right] 
\end{equation}
We adopt the following values for the free parameters: $k_0 = 0.7\,h\,{\rm Mpc}^{-1}$ and $\Delta k = 0.1\,h\,{\rm Mpc}^{-1}$ and check that our results are unchanged if we vary these parameters by 50\%. We then add the low-pass filtered velocities sampled at the location of the DM particles to the displacement field. Our approach for handling redshift-space displacements is not unique, and other methods have been proposed to model the redshift-space clustering of large-scale structure tracers using the HEFT model \citep[see e.g.,][for an application on galaxies]{2023MNRAS.520.3725P}. A nice property of the field-level HEFT model is that in principle it can be used to model not only the two-point correlation function (power spectrum in Fourier space), but also higher-order statistics such as the 21 cm bispectrum, which are expected to contain much of the information, since the 21 cm field is highly non-Gaussian \citep{2022MNRAS.510.3838W}.

We summarize the empirical steps we employ in order to obtain the advected operators as follows:
\begin{enumerate}
    \item We use the initial conditions at $z_{\rm IC} = 49$ to calculate the 4 fields $(\delta_L, \delta_L^2, s_L^2, \nabla^2 \delta_L)$ on a cubic grid of size $128^3$ in Fourier space. We note that we have tested our entire pipeline at a resolution of $1024^3$ and find that our results are unchanged at the scales of interest.
    \item We then evolve the initial condition fields linearly to some redshift of choice, $z$, by tagging each particle at $z$ by the values of the four initial conditions fields and then performing TSC interpolation, using these tags as weights. Including a case in which we set the weights to one, this yields the following five advected bias operators: $(1^{\rm adv.}, \delta_L^{\rm adv.}, \delta_L^{2, \ \rm adv.}, s_L^{2, \ \rm adv.}, \nabla^2 \delta_L^{\rm adv.})$ by assigning each particle a weight given by the value of its position in the initial conditions.
    \item Finally, we store all the advected fields at that redshift $z$ and construct a model of our observable as a linear combination of these operators:
    \begin{eqnarray}
    1 + \delta_{21} &= 1^{\rm adv.} + b_1 \delta_L^{\rm adv.} + b_2 \delta_L^{2, \ \rm adv.} \\ \nonumber
    &+ b_s s_L^{2, \ \rm adv.} + b_{\nabla} \nabla^2 \delta_L^{\rm adv.} .
    \end{eqnarray}
\end{enumerate}
We note that our expansion is performed in the product of the neutral hydrogen fraction and the gas density rather than as a combined expansion in both fields separately, which is more natural from the perspective of perturbation theory applied to large-scale structure \cite{2022PhRvD.106l3506Q,2022JCAP...10..007S}.

\subsection{Hybrid Lagrangian operators}

\begin{figure}
\includegraphics[width=\columnwidth]{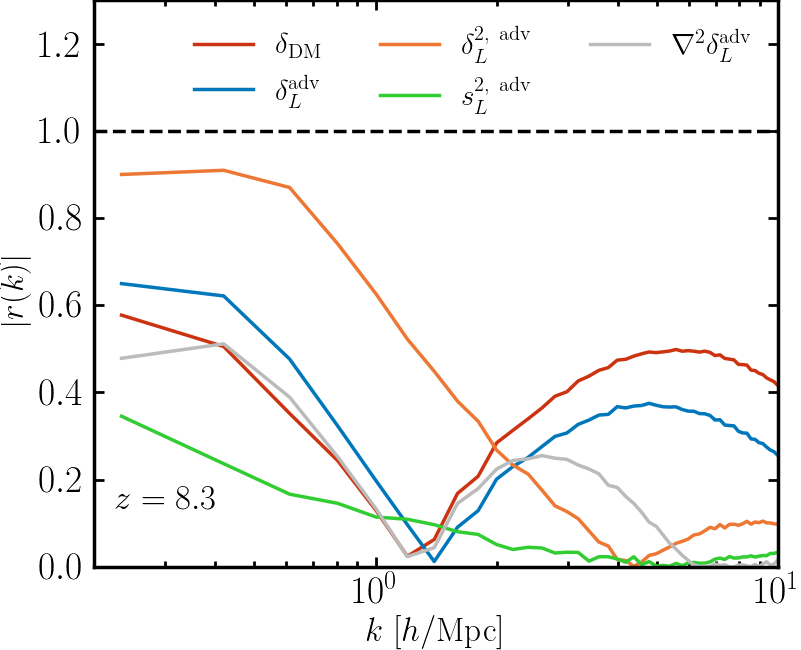}
\caption{Cross-correlation coefficient, $r$, between the redshift-space 21 cm field and the matter field and advected fields at $z = 8.3$. The correlation with the DM field (red line) peaks at 60\% on the largest available scales for our box, $k \sim 0.1 \hompc$ then crosses zero around $k \approx 1 \hompc$, due to bubble formation, and finally peaks on very small scales $k \approx 6 \hompc$, where the Universe is mostly ionized and only small pockets of neutral hydrogen survive. Interestingly, the $\delta_L^{\rm 2, \ adv.}$ operator exhibits the strongest correlation ($r \gtrsim 0.9$) on large scales and then drops steadily on small scales. At this redshift, $s^{\rm 2, \ adv.}_L$ is the least correlated operator on large scales. We note that to create this plot, we use a finer grid size than in the rest of the paper $\sim$60 $\mpcoh$ for all operators except for $\nabla^2 \delta_L^{\rm adv.}$, for which we opt for a grid size of 0.5 $\mpcoh$ due to its numerical instability.}
\label{fig:rk_fields}
\end{figure}

The end product of our theoretical model for the 21 cm field is the Lagrangian fields, $\delta_L$, $\delta^2_L$, $s^2_L$ and $\nabla^2 \delta_L$, advected to some redshift of interest using an $N$-body simulation. In this section, we compare their power spectra to a theoretical prediction coming from the second-order LPT code, \texttt{velocileptors} \cite{2020JCAP...07..062C} and study how strongly correlated these fields are with our observable, the 21 cm field.

We first visualize the advected fields (Lagrangian operators) as well as the 21 cm field in Fig.~\ref{fig:fields}. These are shown at $z = 9$ when the Universe is 80\% neutral and the bubbles have not yet started to coalesce into large ionized regions. Visible in this plot are the bubble nucleation sites, the sharp ionization fronts where ionizing photons get absorbed, and the regions of high density in neutral hydrogen. 

Looking at the higher-order Lagrangian operators, we notice that the $\delta_L^{2, \ {\rm adv.}}$ field highlights the regions with high bias, the $s_L^{2, \ {\rm adv.}}$ emphasizes regions with large anisotropies, and $\nabla^2 \delta_L^{{\rm adv.}}$ accentuates the smallest scales aiming to capture non-local effects. We display our observable as well as the Lagrangian operators at the fiducial resolution of this analysis, $N_{\rm mesh} = 128^3$, which corresponds to a cell size of $0.5 \mpcoh$. This is sufficiently fine to resolve the scales of cosmological interest for future radio interferometer experiments, i.e.\ Fourier modes smaller than $k \lesssim 1 \hompc$, or configuration-space scales larger than $r \sim 5 \mpcoh$.

We next compare the 15 combinations of cross-power spectra constructed from our five numerically advected fields against the predictions from second-order LPT obtained using the \texttt{python} package \texttt{velocileptors}\footnote{\url{https://github.com/sfschen/velocileptors}}. We compute the power spectrum using 20 bins between $k_{\rm min} = 0$ and $k_{\rm max} = 1 \hompc$, and note that the fundamental mode of the box is $2 \pi/L_{\rm box} \approx 0.1 \hompc$. We show this comparison at $z = 9$ in Fig.~\ref{fig:comparison}, but we find a consistent picture for all other redshifts used in this study.

We see good agreement overall between theory and numerics: terms involving the  $\delta_L^{\rm adv.}$ and $s_L^{\rm 2, \ adv.}$ tend to agree best with theory, whereas terms involving the  $\nabla^2 \delta_L^{\rm adv.}$ and $\delta_L^{\rm 2, \ adv.}$ exhibit the largest deviations from theory likely because both are particularly sensitive to small-scale noise due to aliasing effects. The numerical noise is particularly pronounced on all scales for the $\delta$-squared operators, $\delta_L^{\rm 2, \ adv.}$ and $s_L^{\rm 2, \ adv.}$. We note that smoothing of the initial conditions fields before performing the advection helps in reconciling these differences. Thus, the cause of the differences in all of these operators is well-understood from a theoretical point of view and irrelevant to the analysis performed in this work, as it can be diminished via a renormalization.

We are also interested in understanding how correlated the advected fields are with our observable of interest, the 21 cm field $\delta_{21}$. To this end, we compute the cross-correlation coefficient between $\delta_{21}$ and the advected fields, which allows us to quantify the amount of correlation, as follows: 
\begin{equation}
    r^{AB}(k) = \frac{P^{AB}(k)}{\sqrt{P^{AA}(k) P^{BB}(k)}} .
\end{equation}
Qualitatively, we expect the $r(k)$ coefficient to drop sharply on small scales once we approach the typical scale of bubbles at the given redshift, which corresponds to the scale of non-locality.  In Fig.~\ref{fig:rk_fields}, we show this quantity at $z = 8.3$ for all five advected fields. We note that qualitatively this plot looks very similar at the higher and lower redshifts we consider, with the higher redshifts yielding generally higher correlation and the lower redshifts generating a lower correlation. This can be easily understood as at these higher redshifts there are fewer bubbles to decorrelate the signal from the underlying density fields as well as the Universe is more linear as structure has barely begun to collapse gravitationally.

While on large scales, $k \sim 0.1 \hompc$, we are limited by cosmic variance due to the small volume of the simulation, recently a powerful technique known as Zel'dovich Control Variates has been developed and applied to cosmological observables such as the power spectrum and the correlation function to significantly reduce the sample variance of simulation-measured quantities \citep{2022JCAP...09..059K,2023JCAP...02..008D,2023OJAp....6E..38H}. This method performs best in the regime where the cross-correlation coefficient $r(k)$ between the observable of interest and the Lagrangian/Zel'dovich operators is close to one, as is the case for the 21 cm field at $z \gtrsim 7$ (i.e., when the bubble size is smaller than the box, and $r(k)$ is therefore close to one). Since radiative transfer hydrodynamical simulations are prohibitively expensive to run in Gpc-sized boxes, the Control Variates technique could be extremely useful for partially overcoming the volume limitation. 

Looking at Fig.~\ref{fig:rk_fields}, we see that the correlation with the DM field peaks at 60\% on the largest available scales for our box, $k \sim 0.1 \hompc$ then crosses zero around $k \approx 1 \hompc$, the characteristic bubble size at this redshift, and finally peaks at very small scales $k \approx 6 \hompc$. We note that on these scales the Universe is mostly ionized and only small pockets of neutral hydrogen exist due to recombination and self-shielding. Interestingly, these surviving pockets of neutral hydrogen are traced well by the regions of high density. Furthermore, we see that the $\delta_L^{\rm 2, \ adv.}$ operator exhibits the strongest correlation ($r \gtrsim 0.9$) on large scales and then drops steadily on small scales, crossing zero at $k \approx 4 \hompc$, which occurs on smaller scales than the bubble size due to the mode-coupling of Fourier modes. The $s^{\rm 2, \ adv.}_L$ is the least correlated operator on large scales. As will be discussed later, at these redshifts $z \simeq 8-9$, the linear bias appears to vanish, i.e.\ $b_1 = -1$ in the Lagrangian picture (see Fig.~\ref{fig:bias}), which implies that the higher-order operators dominate, which provides a plausible explanation for the high cross-correlation coefficient between the $\delta^{2, {\rm adv.}}_L$ field and $\delta_{\rm 21}$.

We interpret this due to the fact that the matter displacements are much smaller than the distances photons travel during reionization (i.e., the bubble sizes). In addition, because the sources driving reionization are highly biased, we expect their tidal bias, which is proportional to the 21 cm tidal bias, to be small \citep{2018JCAP...10..016M}. Finally, the $\nabla^2 \delta_L^{\rm adv.}$ behaves similarly to the $\delta_L^{\rm adv.}$ term. This is because $\nabla^2 \delta_L^{\rm adv.}$ can be approximated to first order as $k^2 \delta_L^{\rm adv.}$. Thus, the bias parameter associated with that field can be linked to the characteristic bubble size $R_\ast$, as $b_\nabla \sim R_\ast^{1/2} \sim k_\ast^{-1/2}$, where $k_\ast$ is the scale at which this term surges up (and $r(k)$ reaches zero). To create this plot, we use a finer grid size than used in the rest of the paper: $\sim$60 $\mpcoh$ for all operators except for $\nabla^2 \delta_L^{\rm adv.}$, for which we adopt a grid size of 0.5 $\mpcoh$ due to its numerical instability on small scales.

\section{Results}
\label{sec:results}

\begin{figure}
\includegraphics[width=\columnwidth]{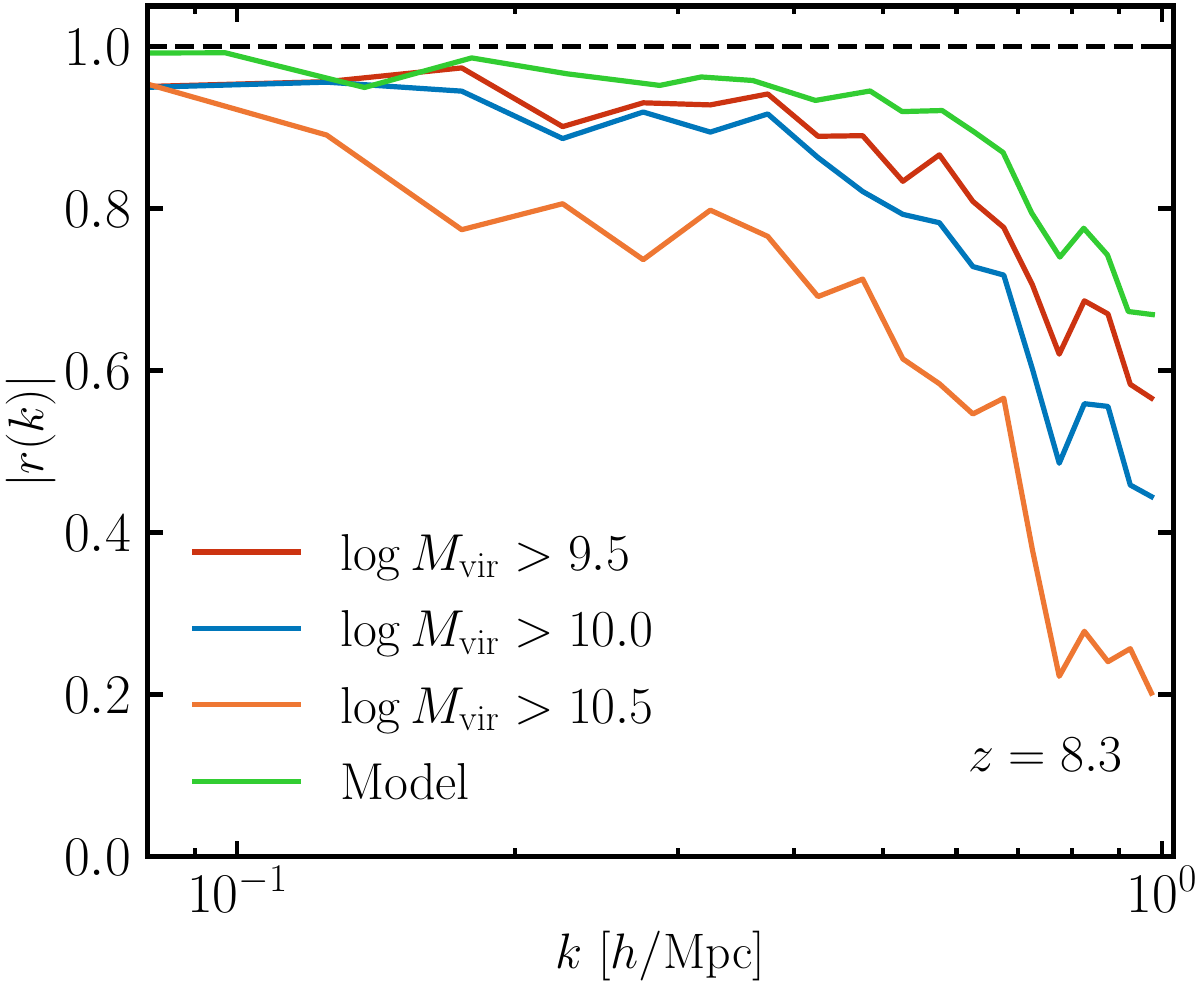}
\caption{Cross-correlation coefficient between halos and the 21 cm field. We split the halos by virial mass into three samples: $\log M_{\rm vir} > 9.5, \ 10, \ 10.5$, where the masses are expressed in units of $h^{-1} M_\odot$. Since on large scales the 21 cm signal traces large-scale structure, we expect these fields to be strongly correlated. The cross-correlation is highest for the lowest-mass sample, as it has the lowest bias, i.e. is closest to the matter field that the neutral hydrogen traces, and has the highest number density, i.e. the lowest level of shotnoise, which additionally decorrelates the two fields.}
\label{fig:rk_halos}
\end{figure}

In this section, we present the main findings of this study. We first describe our bias-fitting method and then assess its performance in terms of the cross-correlation coefficient between our model and the 21 cm field and the error power spectrum. To put our findings into context, we also draw a comparison between the thermal noise of a futuristic radio interferometer experiment and the theoretical noise of our model.

\subsection{Fitting method}
\label{sec:fit}

Here we elaborate on the methodology to extract and minimize the stochastic contributions in LSS models by enhancing the accuracy of bias parameter estimates \citep[see][for more details]{PhysRevD.100.043514,2022MNRAS.514.2198K}. The goal is to estimate the bias parameters \(b_i\) of our model by minimizing the difference 
between the model and the observable $\delta_{\rm 21}$. Our result can be expressed through the stochastic field $\epsilon(\mathbf{k})$, which represents residuals after removing deterministic contributions 
from the model:
\begin{equation}
\epsilon(\mathbf{k}) = \delta_{\rm 21}(\mathbf{k}) - \delta_m(\mathbf{k}) - \sum_i b_i(\mathbf{k}) \mathcal{O}_i(\mathbf{k}),
\end{equation}
We obtain the bias parameters by solving the least squares problem of minimizing the error 
power spectrum, which leads us to the following expression for the bias transfer functions:
\begin{equation}
\hat{b}_i(\mathbf{k}) = \left\langle \mathcal{O}_i \mathcal{O}_j \right\rangle^{-1} (\mathbf{k}) \left\langle \mathcal{O}_j(-\mathbf{k}) \left[ \delta_h(\mathbf{k}) - \delta_m(\mathbf{k}) \right] \right\rangle .
\end{equation}
In particular, we do this as follows:
\begin{equation}
\hat{b}_i(\mathbf{k}) = M_{ij}^{-1}(\mathbf{k}) A_j(\mathbf{k}).
\end{equation}
where $A_j(\mathbf{k})$ and $M_{ij}(\mathbf{k})$ are defined as

\begin{equation}
A_j(\mathbf{k}) = \left\langle [\mathcal{O}_j(\mathbf{x}) (\delta_h(\mathbf{x}) - \delta_m(\mathbf{x}))]_{k_{\text{b}},k_{\text{b+1}}} \right\rangle ,
\end{equation}
\begin{equation}
= \int_{k_{\text{b}} < {|\mathbf{k}| < k_{\text{b+1}}}} 
\frac{d^3k}{(2\pi)^3} \mathcal{O}_j(\mathbf{k}) [\delta_h - \delta_m]^*(\mathbf{k}) ,
\end{equation}
and 
\begin{equation}
M_{ij}(\mathbf{k}) = \left\langle [\mathcal{O}_i(\mathbf{x}) \mathcal{O}_j(\mathbf{x})]_{k_{\text{b}},k_{\text{b+1}}} \right\rangle ,
\end{equation}
\begin{equation}
= \int_{k_{\text{b}} < {|\mathbf{k}| < k_{\text{b+1}}}} \frac{d^3k}{(2\pi)^3} \mathcal{O}_i(\mathbf{k}) \mathcal{O}_j^*(\mathbf{k}), 
\end{equation}
where the subscript b represents the index that runs from 1 to $N_{\rm b}$, which corresponds to the number of bins into which we have divided our Fourier modes. Once the $\hat{b}_i(\mathbf{k})$ coefficients are obtained, they can be substituted back into our model to estimate the error power spectrum as follows:
\begin{equation}
P_{\text{err}}(\mathbf{k}) \equiv \left\langle \epsilon(\mathbf{k}) \epsilon(-\mathbf{k}) \right\rangle .
\end{equation}

While typically in cosmological analyses we marginalize over the bias parameters, they can also be linked to astrophysical quantities relevant to reionization. Following Ref.~\citep{2018JCAP...10..016M}, we can draw the following connections (though we note that our analysis is performed in Lagrangian space):
\begin{itemize}
    \item $b_1$ constrains the source bias and the global neutral fraction and within a simple model is roughly 
    $b_1 \sim - (1-x_{\rm HI}) (2 + b_1^s)$,
    where $b_1^s$ is the source bias. If we assume that the source bias varies more slowly than $x_{\rm HI}$, then we roughly expect $b_1$ to increase (in absolute magnitude), as we plunge deeper into the EoR.
    \item Because reionization is a patchy process and highly correlated with regions of high density (and thus large source bias), the $b_2$ parameter is expected to be quite large and in fact be the dominant term when $b_1 \approx 0$.  
    \item $b_s$ is linked to $b_s^s$ of the source and is likely subdominant, as the scale of displacements in the matter field are smaller than the reionization front scale. 
    \item Finally, the $b_\nabla$ bias associated with the $\nabla^2 \delta_L^{\rm adv.}$ tells us how significant and what scale non-localities occur, which in the case of reionization is related to the characteristic bubble size.
\end{itemize}
At the end of reionization ($z \sim 6$), there is some residual neutral hydrogen found in galaxies,
and thus the 21 cm field effectively becomes a positively biased tracer of large-scale structure. In this regime, perturbative bias expansion methods have been extensively tested and can be straightforwardly applied to model the late-time 21 cm field \cite{2018arXiv181009572C,2019JCAP...09..024M,2019JCAP...11..023M,2021JCAP...12..049S,2023PhRvD.108h3528O,2024arXiv240518559F}.

\subsection{Field-level fits}

\begin{figure}
\includegraphics[width=\columnwidth]{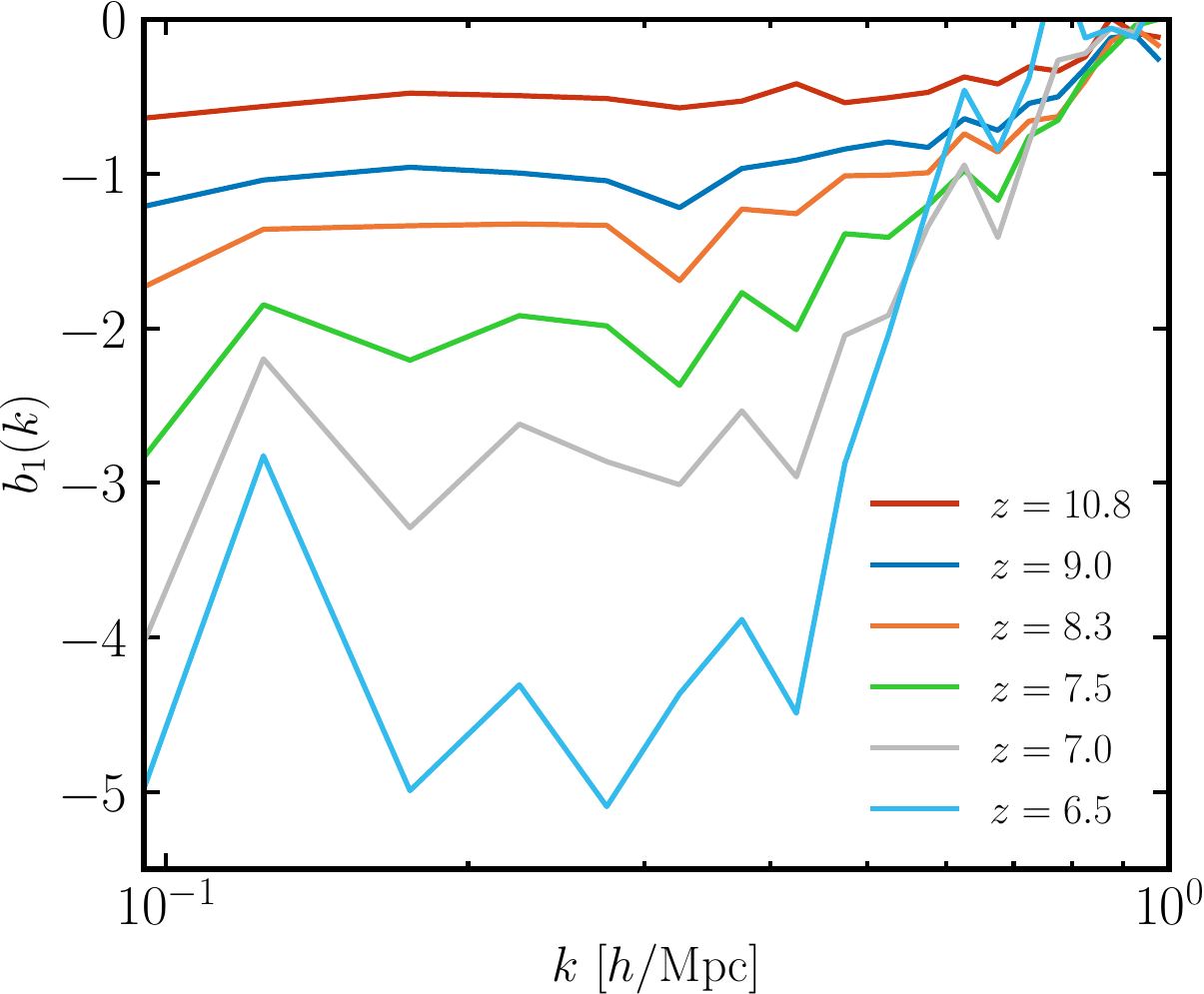}
\caption{Scale-dependent bias parameter, $b_1(k)$, multiplying the first-order Lagrangian operator $\delta_L^{\rm adv.}$ in our expansion model outlined in Section~\ref{sec:theory}. As expected, $b_1$, is stable on large scales, $k \lesssim 0.4 \hompc$,
and then becomes scale-dependent on smaller scales where the quadratic bias model starts to break down as we reach the scale corresponding to the typical bubble size.}
\label{fig:bias}
\end{figure}

We first explore the scale dependence of the first-order bias parameter $b_1(k)$ in Fig.~\ref{fig:bias}, which as argued in Section~\ref{sec:fit} can be connected to the mean reionization fraction and the source bias. As claimed above, in the case of a nearly constant (in time) source bias, we expect $b_1$ to become more and more negative as we go to lower redshifts. This is also what we find in this figure, though we note that these arguments break down towards the end of reionization where the 21 cm field becomes highly anisotropic and non-linear and the bubbles merge to form large ionized regions. We see that as a function of scale, $b_1$ is stable on scales where $k \lesssim 0.4 \hompc$, corresponding to scales larger than the bubble sizes. This characteristic wave number is larger at high redshifts ($k \lesssim 0.8 \hompc$ at $z = 10.8$) and smaller at low redshifts, at which the neutral regions occupy an increasingly smaller volume, making both the measurement more noisy and the theory more inaccurate.

\begin{figure}
\includegraphics[width=\columnwidth]{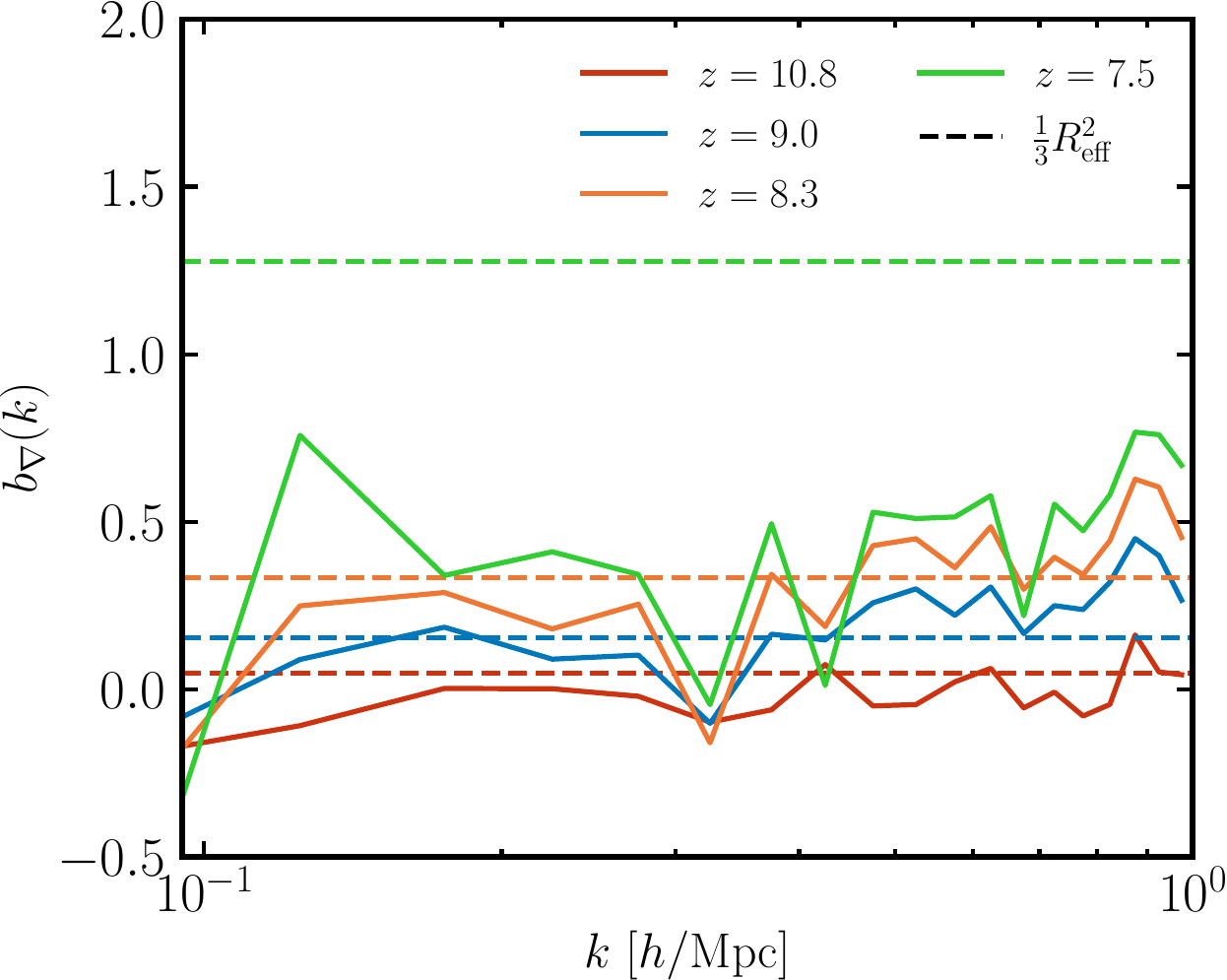}
\caption{Comparison between the theoretical prediction for $b_\nabla \sim 1/3 R_{\rm eff}^2$ (see Section~\ref{sec:fit}), and its numerical value from the scale-dependent fits to the Thesan-1 21 cm signal. We find that at high redshifts, this approximation works pretty well, but at low redshifts ($z \lesssim 7.5$), where the bubble size grows exponentially (see Fig.~\ref{fig:mean_ion}), it starts to break down.}
\label{fig:bias_nabla}
\end{figure}

In Fig.~\ref{fig:bias_nabla}, we explore the connection between the characteristic bubble size, $R_{\rm eff}$, and $b_\nabla$, which as conjectured in Section~\ref{sec:fit} is related to $R_{\rm eff}$ as $b_\nabla \sim 1/3 R_{\rm eff}^2$ (note that $b_\nabla$ is thus expected to be positive). We find that $b_\nabla$ is fairly scale independent at high redshifts, where the approximation works quite well. At low redshifts ($z \lesssim 7.5$), it starts to break down likely because the $k^2 \delta_L$ term struggles to catch up when the bubble size grows exponentially (see Fig.~\ref{fig:mean_ion}).

\begin{figure}
\includegraphics[width=.98\columnwidth]{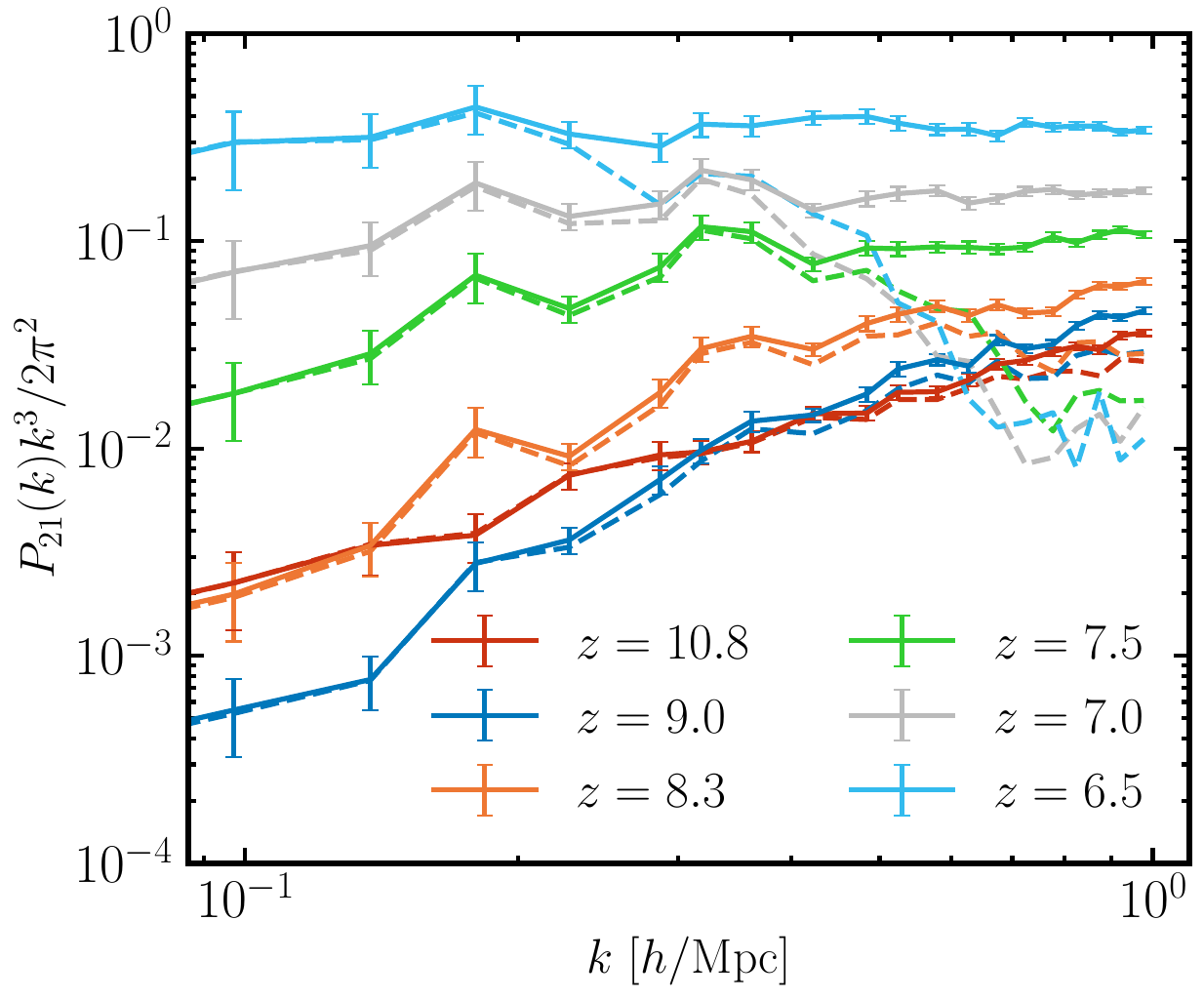}\\ 
\includegraphics[width=.98
\columnwidth]{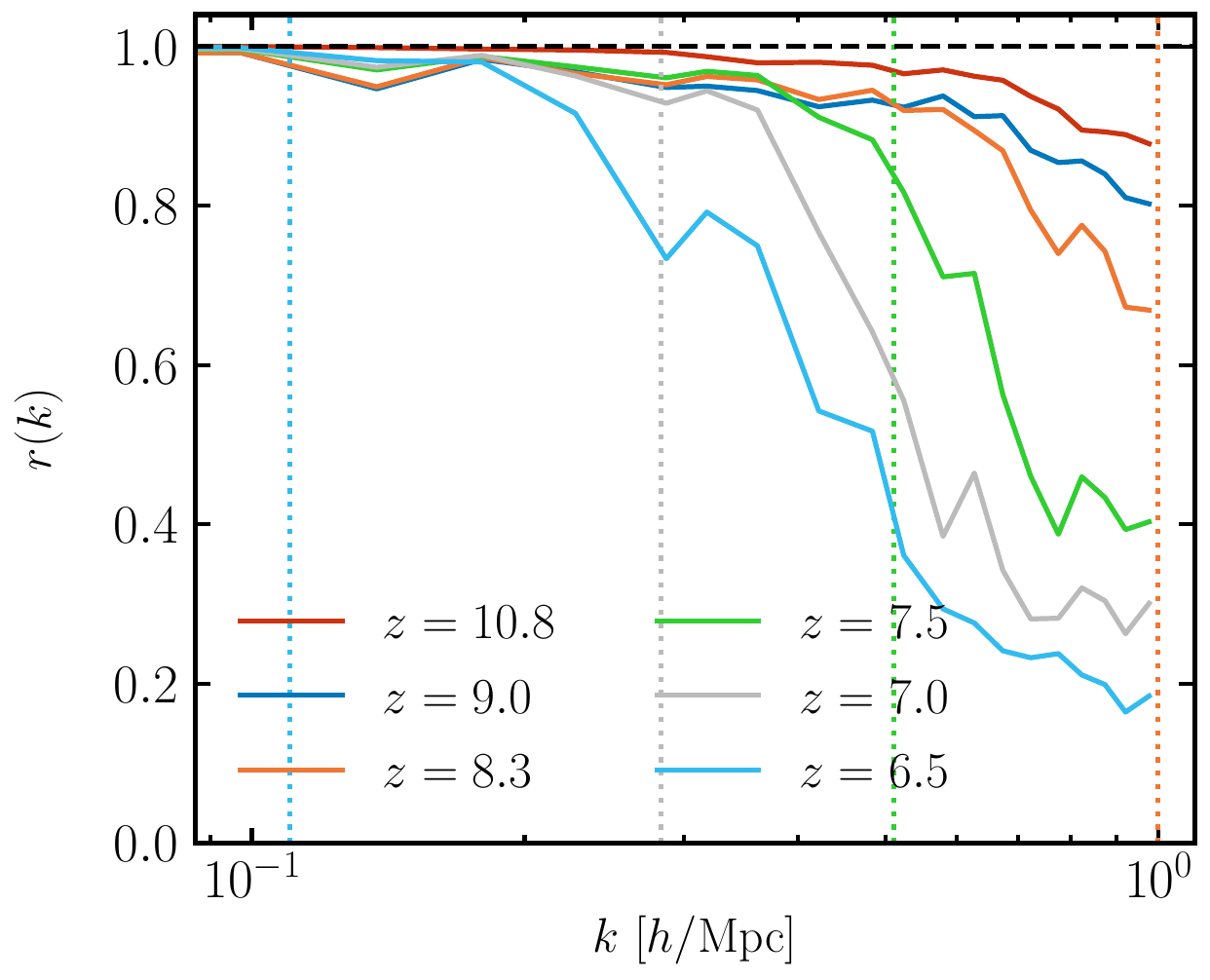}
\caption{Monopole of the redshift-space power spectrum (top) and cross-correlation coefficient (bottom) of the 21 cm field and our model, described in Section~\ref{sec:theory}. Top panel: overall the amplitude of the power spectrum increases on large scales with time as the fluctuations in the matter field continue to evolve linearly. At the same time, bubble formation also causes the power spectrum to drop on large scales, explaining the behavior of the $z = 10.8$ curve relative to the rest. Visually, our model (dashed line) provides a good description of the true 21 cm power spectrum down to small scales, $k \sim 0.6 \hompc$, at high redshifts and starts to deviate from the truth earlier at low redshifts. Bottom panel: consequently, we see this reflected in the cross-correlation coefficient. On large scales we see all redshifts displaying an almost perfect correlation $r \approx 1$, with the $z = 10.8$ curve being consistently larger than $r > 0.9$ on the scales shown, indicating that our model is successful at recovering the scales relevant for cosmological analyses. There is also a significant amount of noise in all curves due to the limited volume of the simulation. The vertical dotted lines correspond to $1/R_{\rm eff}$ at each redshift and indicates the scale of non-locality due to bubble growth.}
\label{fig:pk_rk}
\end{figure}

We next study the monopole of the redshift-space power spectrum and the cross-correlation coefficient between model and observable in Fig.~\ref{fig:pk_rk} for the best-fit bias parameters obtained as described in Section~\ref{sec:fit}.
At very high redshifts $z \simeq 12-9$, the 21 cm power spectrum drops on large scales as ionization bubbles start to form and take out regions with high bias. Around $z \sim 9$, the power spectrum amplitude picks up again as the density fluctuations in the 21 cm field start to evolve. At the same time, due to the growing reionization fronts, the 21 cm signal begins to downgrade on small scales, leading to the characteristic flattening of $P_{21}(k) k^3$ \citep[see e.g.,][for a discussion]{2008ApJ...680..962L}. This is exactly what we observe in the top panel. We also see that the perturbative model provides a good description of the true 21 cm power spectrum on large scales and breaks down as we reach the scales of non-locality, where the model power spectrum becomes deficient. As expected, our model is more successful at high redshifts compared with low redshifts. 

Luckily, the cosmological information to be gained from these lower-redshift slices is likely quite negligible. One can understand this by noticing that the 21 cm power spectrum, $P_{\rm 21}$, is very flat towards the end of reionization (see Fig.~\ref{fig:pk_rk}), so the power spectrum is effectively `featureless' and all we can extract is an amplitude \citep{2008ApJ...680..962L}. This is also made worse by the fact that current and near-future experiments will only measure the signal in a few bandpowers, thus additionally smearing any features that could otherwise yield information. A lot of the power found on small scales is likely due to non-linear astrophysical effects related to the growing and merging population of ionization bubbles. It is thus uncorrelated with the large-scale density field and effectively contributes `noise' from the perspective of our perturbation model. We verify that indeed at $z \approx 6.5$ the correlation between the dark matter field and the 21 cm signal drops off sharply (and more so than at higher redshifts) at $k \gtrsim 0.2 h {\rm Mpc}^{-1}$, providing evidence in support of this conjecture. For completeness, we provide the wavenumber values, $k_{r = 0.5}$, at which the cross-correlation coefficient with the DM field, $r(k)$, drops below 0.5: $k_{r = 0.5} \approx 0.4 \, h^{-1} {\rm Mpc}$ at $z = 6.5$, $k_{r = 0.5} \approx 0.6 \, h^{-1} {\rm Mpc}$ at $z = 7.5$, $k_{r = 0.5} \approx 2 \, h^{-1} {\rm Mpc}$ at $z = 10.8$, and note that $k_{r = 0.5}$ increases smoothly with redshift.

The cross-correlation coefficient between our model and the $21\,$cm field exhibits a similar trend (Fig.~\ref{fig:pk_rk}). On large scales, we see that all redshifts display an almost perfect correlation $r \approx 1$, with the $z = 10.8$ curve being consistently larger than $r > 0.9$ all the way until $k \sim 1 \hompc$. Even at low redshifts, close to the end of reionization, the cross-correlation coefficient is still very close to one up until $k \sim 0.3 \hompc$. This indicates that our model does not simply match the power spectrum out of shear luck but does so at the field level, matching the phases of the 21 cm Fourier modes.  This is important, because the bias parameters affect only the amplitudes of the terms, not their phases. There is a significant amount of noise in all curves, which we attribute to the limited volume of the simulation. We expect $r(k)$ to be more stable and consistently closer to one in a larger-box simulation.
We note that when $r$ becomes small, $r(k) < 1$, then we can say that the 21 cm field is not faithfully tracing large-scale structure in the Universe. 

In addition to the lower cross-correlation coefficient with the matter field, we also verify that this holds for its cross-correlation coefficient with halos (and galaxies, by extension), as expected, since they are biased tracers of large-scale structure. This needs to be taken into account for science cases that rely on cross-correlation of $21\,$cm surveys with surveys of galaxies or star-formation-based line emission. We show the cross-correlation coefficient between halos and the 21 cm field at $z = 8.3$ in Fig.~\ref{fig:rk_halos}. We split the halos by virial mass into three samples: $\log M_{\rm vir} > 9.5, \ 10, \ 10.5$, where the masses are expressed in units of $h^{-1} M_\odot$, corresponding to number densities of $n_{\rm halo} \approx 5 \times 10^{-2}$, $7 \times 10^{-3}$, and $7 \times 10^{-4} h^3 {\rm Mpc}^{-3}$, respectively. The cross-correlation is highest for the lowest-mass sample, as it has the lowest bias, i.e. is closest to the matter field that the neutral hydrogen traces, and has the highest number density, i.e. the lowest level of shotnoise, which additionally decorrelates the two fields. Additionally, the $10^{9.5} \ h^{-1} M_\odot$ threshold corresponds to roughly a virial temperature of $10^{4.1} \ {\rm K}$, which is close to the expected virial temperature of $10^{4} \ {\rm K}$ to form ionizing bubbles \citep{2022GReGr..54..102C}. An open question in EoR studies is the connection between the properties of the ionizing halos and the characteristic bubble size. As proposed in Ref.~\citep{2005MNRAS.363.1031F}, the mean free path of ionizing photons and thus the characteristic bubble size are likely related to the number density and mass of the photon sinks. It is thus through cross-correlations between the 21 cm emission signal and large-scale structure tracers such as galaxies and quasars that we can reveal information about the UV radiation sources such as their characteristic mass. 

\begin{figure}
\includegraphics[width=\columnwidth]{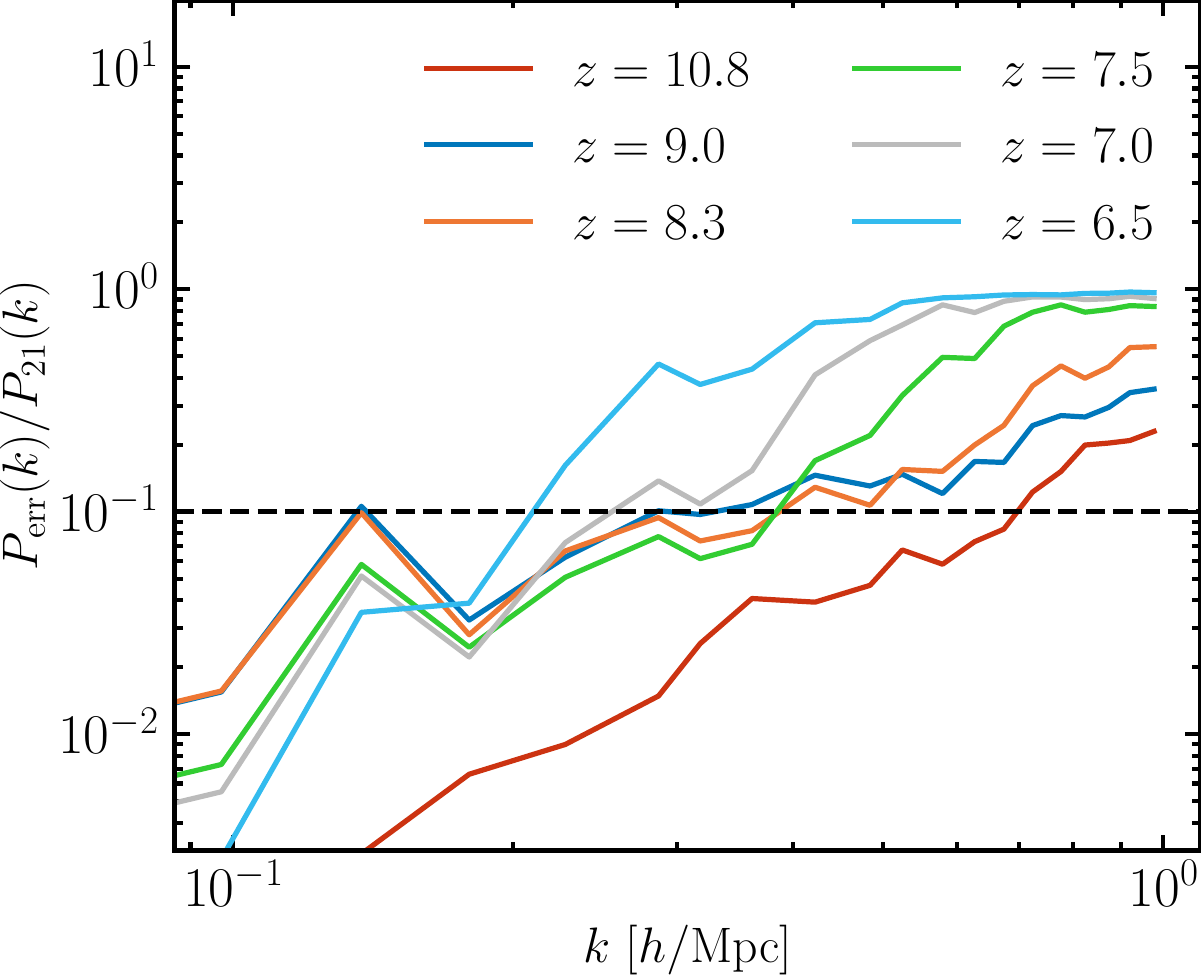}
\caption{Ratio between the error power spectrum (i.e., of the difference between model and observable) and the power spectrum of our observable, the redshift-space 21 cm field, averaged over the LOS angle (monopole). The black dashed curve serves to guide the eye and indicate the 10\% error threshold. As expected, the error is smallest on large scales ($\sim$1\%) and increases on small scales, with the highest redshifts exhibiting more perturbative behavior than the lowest ones. Moreover, we note that the error on the power spectrum is behaved well even on small scales, indicating that our method might be well-suited for analysis even on scales beyond $ k \gtrsim 0.5 \hompc$ in contrast with pure perturbative methods. We note that there is a significant amount of noise on large scales due to the limited number of Fourier modes.}
\label{fig:pk_ratio}
\end{figure}

Due to this limitation in the number of Fourier modes on large scales, we also find a significant amount of noise in the ratio between the error power spectrum (of the difference between model and observable) and the observable power spectrum, studied in Fig.~\ref{fig:pk_ratio}. We find that the error is smallest on large scales, as expected, and is around $\sim$1\%. It then rapidly increases as we push to smaller scales near the scale of non-locality at that redshift. Generally, the highest redshifts have the smallest error power spectra, with the $z = 10.8$ curve reaching the 10\% mark only at $k\approx 0.8 \hompc$. Moreover, we note that the error on the power spectrum is well-behaved even on small scales, indicating that our method might be well-suited for analysis even on scales beyond $k \gtrsim 0.5 \hompc$ and overall seems to perform better in that regime than pure perturbative methods \citep{2018JCAP...10..016M,2022PhRvD.106l3506Q,2022JCAP...10..007S}, where the error tends to increase exponentially on these scales. We do not find a strong dependence of this ratio on the angle with respect to the LOS, suggesting that our method for handling redshift-space distortions is adequate. We thus opt to only show the monopole result, as $k$-$\mu$ slices suffer from a significant amount of cosmic variance. 
We note that the breakdown of the bias expansion towards the end of reionization also suggests that the (small) box does not contain a representative sample of bubble morphologies and sizes at that epoch, i.e.\  $k_{\rm eff}$ is approaching $k_{\rm fund} = 2 \pi/L_{\rm box}$ at late times. In fact, as seen in Fig.~\ref{fig:mean_ion}, by $z \approx 6$, the bubbles cover almost the entire volume of the Thesan-1 simulation, i.e.\ $R_{\rm eff} \sim L_{\rm box}$.

\subsection{Survey realism}

In this section, we present a comparison between the thermal noise of a futuristic experiment and the theoretical noise associated with our model. The goal is to qualitatively explore the question of whether our model is sufficiently good at predicting the 21 cm field signal at the level that meets the sensitivity of next-generation experiments. We note that the frequency coverage of existing and planned experiments is sufficient to encompass most of the relevant range of redshifts. In order to probe the regime of $k_\perp\sim 0.1\,h\,\mathrm{Mpc}^{-1}$ at $z\sim 10$, we need hundred-meter baseline interferometers, which next-generation facilities aim to provide. However, at these intermediate scales and high redshifts, we are sky noise dominated, so the exact baseline distribution is of secondary importance, and we are mostly sensitive to the total collecting area and the integration time. The minimum $k_\parallel$ is limited by the ability of the experiment to subtract foregrounds that are many orders of magnitude brighter than the signal. We note that in this analysis, we ignore other sources of contamination from e.g., Galactic and extragalactic foregrounds such as synchrotron radiation and radio point sources as well as non-uniform instrument noise such as cross-talk, confusion from spurious Earth and electronics radio signals, which can have a spatially anisotropic contribution and thus be confused with the 21 cm signal, especially for large incidence angles, an effect referred to as `foreground leakage.'

\begin{figure*}
\includegraphics[width=0.98\columnwidth]{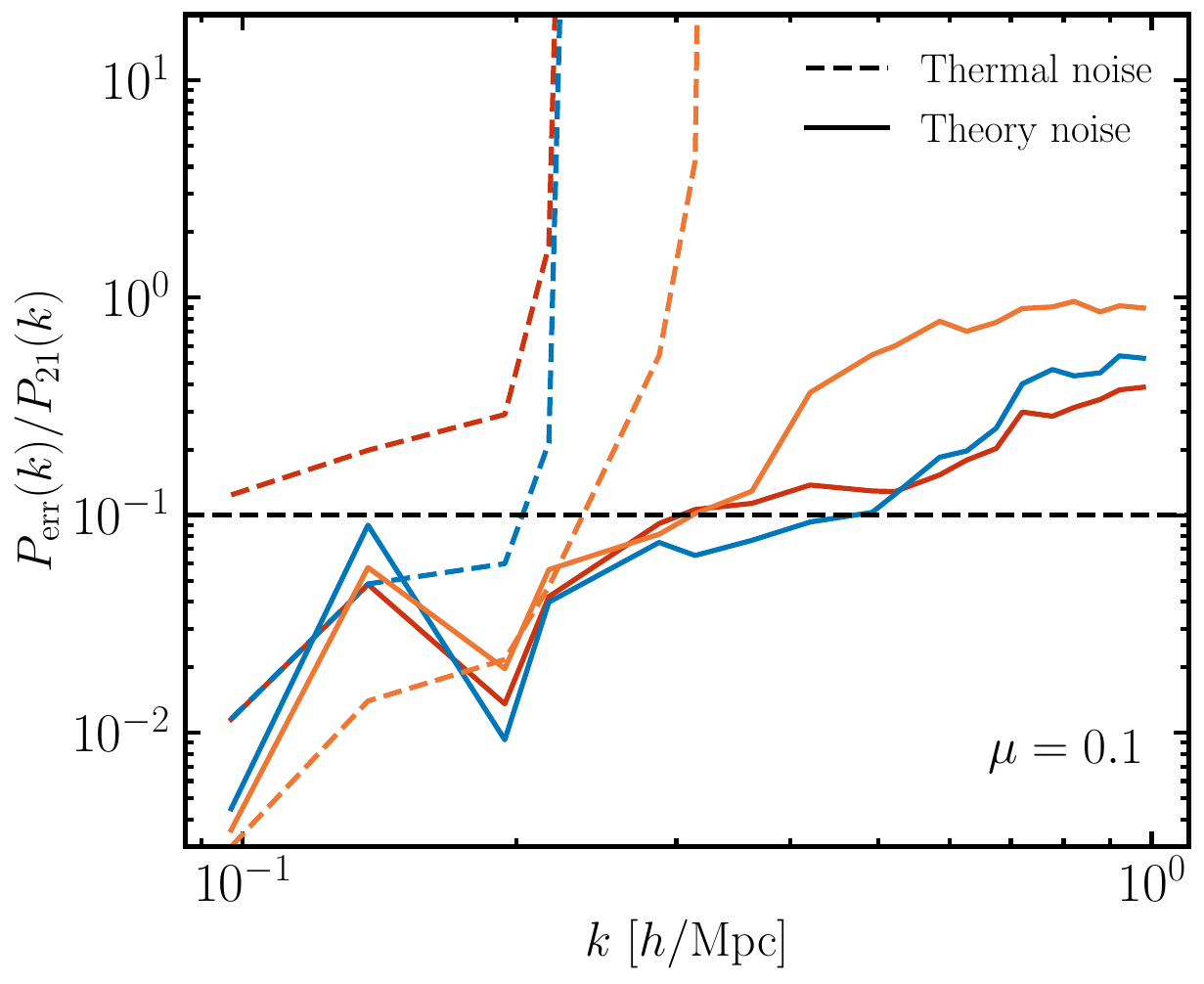}\includegraphics[width=0.98\columnwidth]{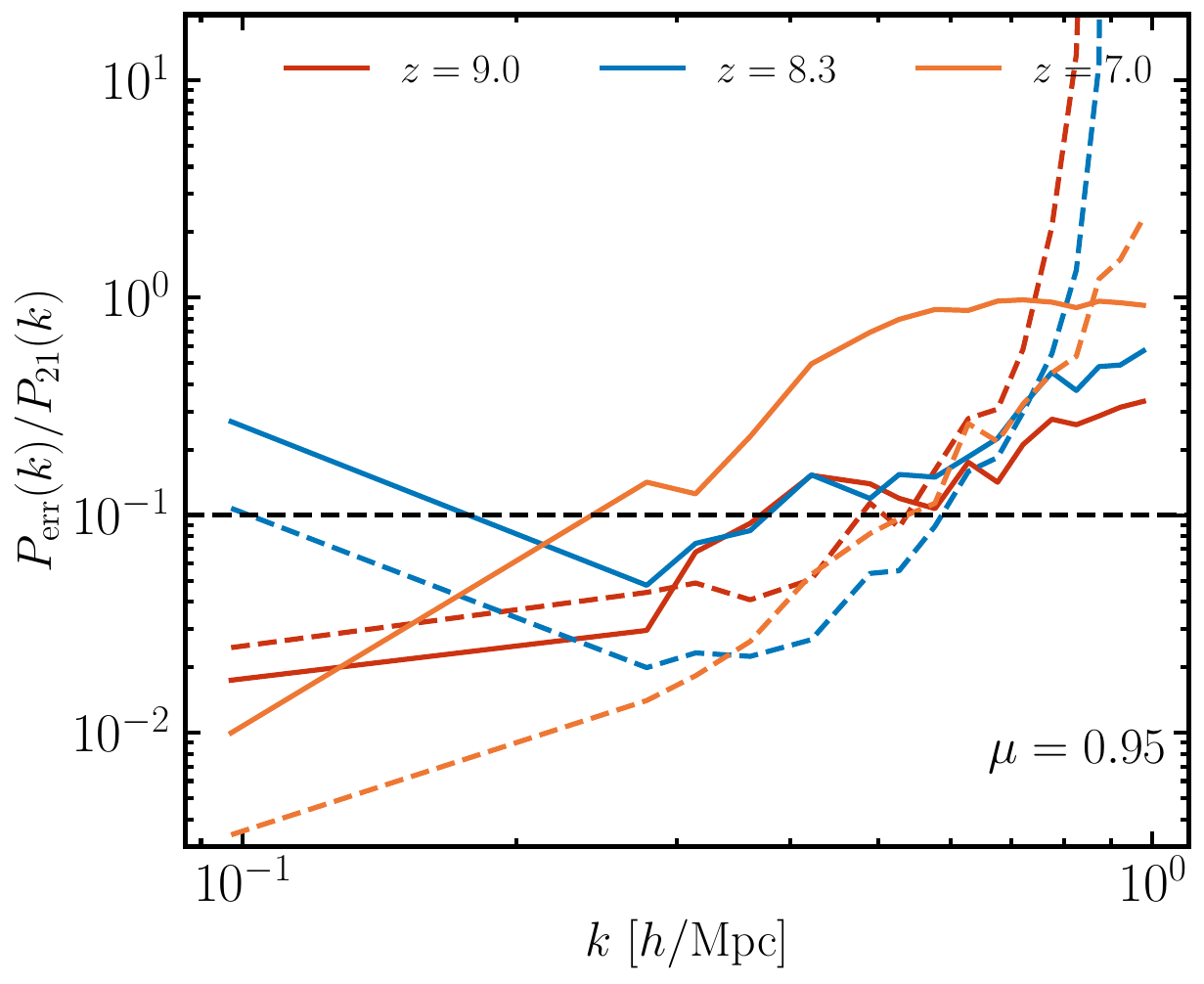}
\caption{Comparison between the error and the thermal noise ratio with respect to the 21 cm power spectrum for the two slices: $\mu = 0.1$ and $\mu = 0.95$, at three different redshifts. It is evident that for the futuristic HERA-like experiment described in this work (and ignoring the presence of foreground leakage and other systematic errors), the thermal noise is comparable with the theory error, indicating that our model provides a viable description of the 21 cm field on the scales accessible by terrestrial radio interferometers.}
\label{fig:pk_ratio_exp}
\end{figure*}


Here, we consider a baseline interferometer instrument that covers 10\% of the sky with a cadence of $\Delta z = 0.2$ at $7 < z < 10$ and a hexagonally packed array of $10^3$ 14 m dishes. This design is in effect a scaled-up version of the Hydrogen Epoch of Reionization Array \citep[HERA][]{2017PASP..129d5001D}. The integration time we consider is 2 years (inversely proportional to the thermal noise), and we adopt a power spectrum binning of $\Delta k = 0.05$, $\Delta \mu = 0.01$. 
In this analysis, we ignore all contributions to the noise budget apart from thermal noise, and we direct the reader to Ref.~\citep{2012ApJ...752..137M} for a more thorough discussion of `foreground wedges'. 
We compute the thermal noise power spectrum following Appendix D (see Eq. D4) of Ref.~\cite{2018arXiv181009572C} (see also Refs.~\cite{2015ApJ...803...21B,2010ApJ...721..164S}) with the baseline distribution calculated with \verb|21cmSense|\footnote{\url{https://github.com/steven-murray/21cmSense}}. We then divide the thermal noise power spectrum by the mean 21cm brightness temperature squared to convert from $({\rm temperature})^2\times{\rm volume}\to{\rm volume}$ units.
We note that because the foregrounds vary smoothly in frequency, most of their power is confined to large-scale modes parallel to the line of sight, so we typically filter out the lowest $k_\parallel$, which removes the bulk of the foreground contamination.

As noted above, we do not explicitly model a foreground `wedge'.  Our understanding of instrument calibration and data analysis techniques for minimizing the influence of the wedge are evolving rapidly, so any choice would be quickly rendered obsolete.  We will note that in the presence of a significant wedge only the high $k$ and high $\mu$ regime would be accessible.  This is unfortunately where the cosmological information is minimized compared to the complex details of the reionization process and where our models perform the worst.  Thus, qualitatively, our ability to extract large-scale structure information depends upon our success in handling the wedge. We direct the reader to Refs.~\citep{2014PhRvD..90b3019L,2015MNRAS.447.1705P,2016MNRAS.456.3142S,2020PASP..132f2001L,2022ApJ...925..221A} for extensive discussions of foreground issues in current and future experiments.

In Fig.~\ref{fig:pk_ratio_exp}, we show the comparison between the error and the thermal noise ratio with respect to the 21 cm power spectrum for the two slices: $\mu = 0.1$ and $\mu = 0.95$, at three different redshifts. It is evident that for the futuristic HERA-like experiment described in this work (and ignoring the presence of foreground leakage and other systematic errors), the thermal noise, which is the absolute minimum noise threshold, is comparable with the error of the model we propose. 
This indicates that the HEFT method developed in this work provides a viable description of the 21 cm field on the scales accessible by terrestrial radio interferometers, and even in idealized conditions (of no foregrounds) can be utilized to analyze the signal with a high degree of success.

\section{Discussion and conclusions}
\label{sec:conclusions}

In this work, we have developed a hybrid effective field theory (HEFT) model that combines $N$-body simulations and a quadratic bias model to describe the redshift-space spatial fluctuations in the 21 cm radiation field during the epoch of reionization. We compare our model predictions to the radiative transfer hydrodynamical simulation Thesan and demonstrate that the 21 cm signal is well-approximated by our HEFT model over the observationally relevant wavenumber range. There exist other methods for modeling the 21 cm signal, including directly through the use of radiative transfer hydrodynamical simulations \citep{2017MNRAS.466..960P,2022MNRAS.514.3857K}, semi-analytic \citep{2022MNRAS.511.3657M}, semi-numerical \citep{2022ApJ...927..186T}, and purely perturbative bias expansion \citep{2019MNRAS.487.3050H,2018JCAP...10..016M,2022PhRvD.106l3506Q} models. Our approach performs a symmetries-based bias expansion in Lagrangian space and advects the initial conditions fields by directly looking up the non-linear displacements in an $N$-body simulation. Our method thus takes advantage of the fact that despite the complexity of the physical processes that govern reionization, the 21 cm field must still obey certain symmetries (e.g. rotational invariance and the equivalence principle), allowing us to construct a model that is valid on scales larger than the typical size of an ionized bubble.

In particular, we study the performance of our model in terms of metrics relevant to cosmological analyses such as power spectrum, cross-correlation coefficient and model error at the field-level. The performance at the power spectrum and correlation function level is summarized in Fig. \ref{fig:pk_rk} for 6 different redshifts, ranging from $z = 6.5$ and 10.8, with corresponding neutral hydrogen fraction of 0.92 and 0.3. Visually, we see excellent cross-correlation on the largest scales available in the Thesan simulation ($k \approx 0.1\,h\,\mathrm{Mpc}^{-1}$), which rapidly falls off on small scales. As expected, at the dawn of reionization, when the bubble sizes are much smaller, the HEFT model works very well. As can be seen in Fig.~\ref{fig:pk_ratio} where we study the error power spectrum ratio, even deep into reionization our model is capable of recovering the 21 cm power spectrum well. We show that our method works best at high redshifts and on large scales, where we find agreement between model and the observable at $\lesssim$ 10\% down to $k \lesssim 0.5 \hompc$, with that wavenumber increasing to $k \approx 0.8 \hompc$ at $z = 10.8$ and dropping down to $k \approx 0.2 \hompc$ at $z = 6.5$. 

\begin{figure}
\includegraphics[width=0.98\columnwidth]{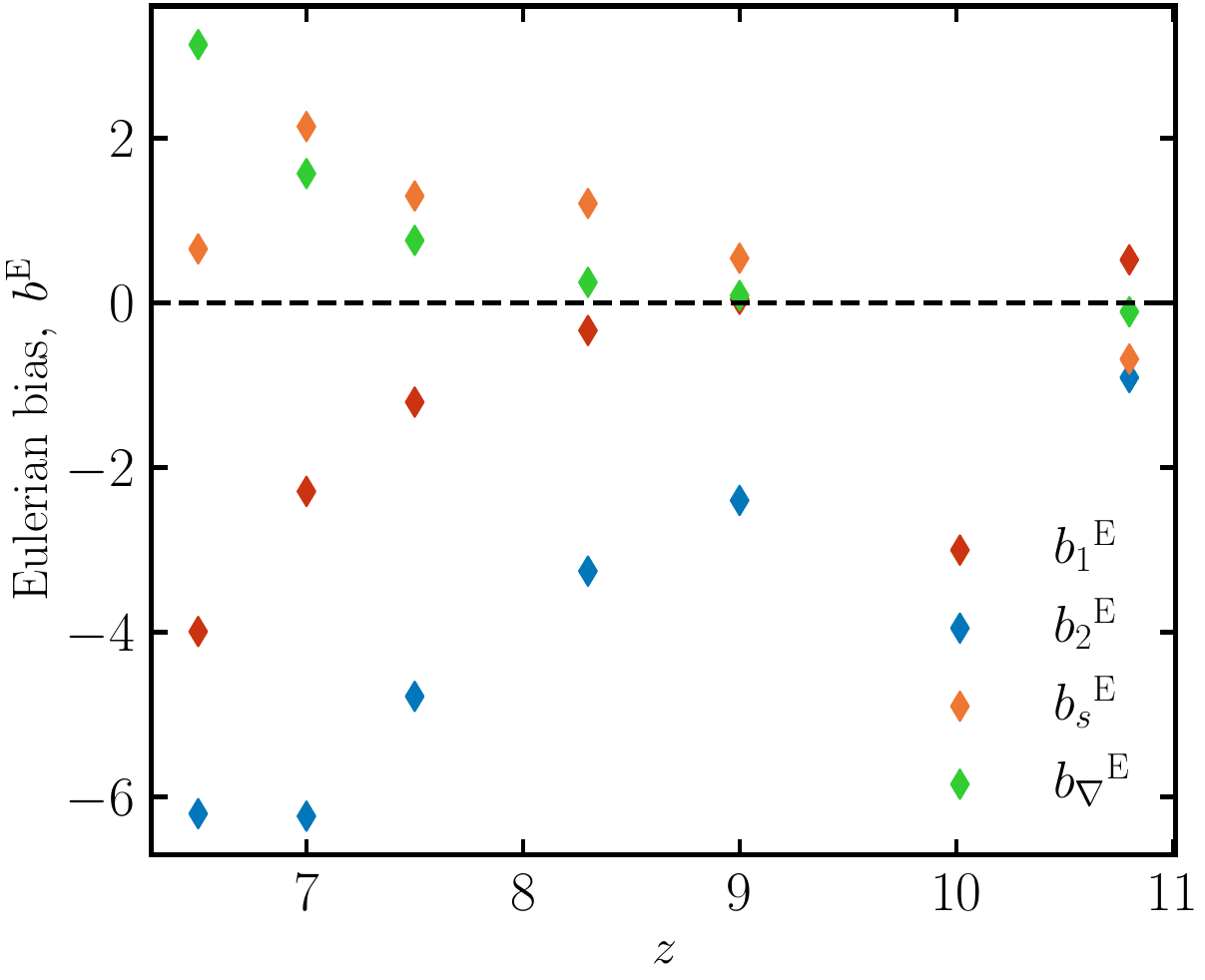}
\caption{Evolution of the Eulerian bias coefficients, obtained by converting our best-fit Lagrangian biases to the Eulerian picture according to App. A of Ref.~\citep{2024arXiv240518559F}. Qualitatively, we find very similar trends to Ref.~\citep{2022PhRvD.106l3506Q} for the parameters $b_1^{\rm E}$, $b_2^{\rm E}$ and $b_s^{\rm E}$, while $b_\nabla^{\rm E}$ grows steadily as the bubble size squared in our model and is exponentially increasing and negative in the model of Ref.~\citep{2022PhRvD.106l3506Q}. Remaining differences may be attributed to renormalization and scale cut choices.}
\label{fig:bias_z}
\end{figure}

We compare our findings for the bias parameter evolution to Ref.~\citep{2022PhRvD.106l3506Q}, which develops a Standard perturbation theory and tests it against the Thesan suite of simulations. In Fig.~\ref{fig:bias_z}, we display the evolution of the Eulerian bias coefficients, obtained by converting our best-fit Lagrangian biases to the Eulerian picture biases according to App.~A of Ref.~\citep{2024arXiv240518559F}. Qualitatively, we find very similar trends to Ref.~\citep{2022PhRvD.106l3506Q}. Namely, $b_1^{\rm E}$ is close to zero at high redshifts and becomes increasingly more negative as reionization progresses. $b_2^{\rm E}$ is negative throughout the EoR and is largest in terms of its absolute value. This also provides a plausible explanation for the high cross-correlation coefficient between $\delta^2_L$ and $\delta_{\rm 21}$ we find in Fig.~\ref{fig:rk_fields}. As expected from previous arguments in Section~\ref{sec:fit}, $b_s^{\rm E}$ is positive and small in magnitude. Specifically, $b_\nabla^{\rm E}$ differs the most: we find that in our model it is proportional to the bubble size squared and increases steadily in magnitude, whereas in Ref.~\citep{2022PhRvD.106l3506Q} it is negative and grows exponentially. We attribute remaining differences to renormalization choices and scale cuts used in the fits. The relatively large values of $b_2$ compared with $b_1$ are not concerning, as the $|\delta^{\rm adv.}|$ field is quite small and $b_2$ multiplies an even smaller quantity, $|\delta^{2, \ {\rm adv.}}|$. We caution that at the lowest redshifts, where the size of the bubbles is comparable to the size of the simulation box, the bias parameters cannot be reliably measured. 

Finally, we study the thermal noise of a futuristic baseline interferometer experiment and compare it with the theoretical noise associated with our model in Fig.~\ref{fig:pk_ratio_exp}. We find that our model performs well in the range of scales that such an experiment would be sensitive to. This suggests that a HEFT-like model is more than suitable to model the observed data in the range of interest.

In the future, we aim to further validate our model in simulations of larger volumes and different resolutions. We conjecture that the photon sinks in high-resolution simulations (which are reflective of photon sinks in the Universe) may be very small which would lead to a decrease in the characteristic size of bubbles relative to Thesan-1 and thus would extend our method deep into the large-wavenumber regime. We also aim to investigate the effect of foregrounds and foreground leakage more thoroughly by e.g., running the `clean' output of Thesan-1 through an end-to-end simulator of a future 21 cm experiment.

Finally, a potential simplification of our model involves using a `cheaper' and purely theoretical forward model that uses e.g. Zel'dovich or second-order LPT displacements as opposed to $N$-body displacements\footnote{See \url{https://github.com/andrejobuljen/Hi-Fi_mocks} for an example of such a model.}, since at most redshifts during reionization, the non-linear scale occurs at larger $k$ modes than the scale at which the bias expansion breaks down due to the reionization bubbles growing. 

\acknowledgements
We would like to thank Aaron Smith, Rahul Kannan, Nick Kokron, Ruediger Pakmor and Wenzer Qin both for several fruitful discussions during the preparation of this paper, and we are especially grateful to Enrico Garaldi and Meredith Neyer for providing additional information for running the initial conditions code and studying the bubble size distribution.

We wish to acknowledge the support of the N3AS undergraduate research program at UC Berkeley for supporting D.B. through the academic 2023-2024 year. B.H. is grateful to the Miller Institute for their generous support.  



\bibliography{refs}{}
\bibliographystyle{prsty}



\end{document}